# The shadow interpretation versus quantum paradoxes

Warren Leffler
Department of Mathematics, Mills College,
5000 McArthur Blvd.
Oakland, California 94613, USA
wleffler@mills.edu

## ABSTRACT

This paper explores the consequences of denying the "emptiness of paths not taken," EPNT, premise of Bernstein, Greenberger, Horne, and Zeilinger (BGHZ) in their paper titled, *Bell theorem without inequalities*.[1] Carrying out the negation of EPNT leads to the concept of a "shadow stream." Streams are essentially particle implementations of the paths in Feynman path-integrals, resulting in a simple and consistent extension of the standard axioms for quantum mechanics. The construct provides elegant resolutions of single- and multi-particle interference paradoxes. Moreover, combining the argument of this paper with that of BGHZ shows that there are just two choices for quantum foundations: interpretations closely similar to the present one or those that harbor instantaneous action at a distance.

## I. INTRODUCTION

The shadow interpretation (SI) presented in this paper derives from a 1993 paper by Bernstein, Greenberger, Horne, and Zeilinger (BGHZ) [1] and also from a 1998 qualitative description by David Deutsch of what he called "shadow particles."[2]

In an important paper for quantum foundations, BGHZ employed a (now) familiar, two-particle interferometer setup [3] to argue that the experimental outcomes for two entangled particles cannot be reproduced by a local, realistic model. Such a model (from the famous EPR paper [4]) forbids interaction between events in space-like separated regions, while also requiring that measurement outcomes reveal pre-existing properties of observables. In arguing against the EPR program and in favor of what they call "Bell theorem without inequalities," BGHZ find it necessary to augment the principles of EPR with the premise, EPNT. EPNT stands for "emptiness of paths not taken." Regarding this they write,

> … one can deny EPNT and thereby imagine that *something* could travel down the empty beam, so as to provide information to the nonempty beam, when the two beams meet. And this something could be consistent with EPR locality, if the particles (and these somethings) on opposite sides of the origin do not communicate.

The shadow interpretation explores the consequences of denying EPNT—that is, what results when the beam is viewed as containing hypothetical "somethings," which we call "shadow particles," a term coined by Deutsch in a different but related context. Deutsch conceived of shadow particles from a different perspective than we do here. For him their origin derives from Hugh Everett's many-worlds interpretation of quantum mechanics (MWI). Deutsch argued that when an ordinary particle (a *tangible* particle) is emitted from a source to, say, a detecting screen in a double-slit experiment, the fringe pattern observed on the screen comes from the interaction of the tangible particle with shadow counterparts coming from parallel universes. The shadow particles interfere only with tangible particles of the same type, and therefore they can be detected only indirectly, through their effects on regular, tangible particles—that is, shadow photons interfere with regular photons, shadow electrons with regular electrons, and so on. In the case of photons we can, as Deutsch explains, narrow the gaps between slits and easily calculate a lower bound of one trillion for the number of shadow particles accompanying any ordinary particle. Now although the shadow interpretation is conceptually related to MWI, and is consistent with it, it differs from MWI in an important way: It is what might be called an "internal model," treating all particle interaction as taking place within a single world, the one we experience. This immediately reduces MWI's "heavy load of metaphysical baggage," as John Wheeler once called it [5] (although it turns out that the argument of Sec. IV renders largely irrelevant such "Occam's-razor" issues).

Whenever a single tangible particle travels between two points, we associate shadow particles with the other *possible* paths between the points, with the particular path taken by the tangible particle being a random event. The collection consisting of the tangible particle along with its shadow counterparts is called a "shadow stream." Richard Feynman's path-integrals [6] (sums) are a perfect match for quantifying the amplitudes of shadow streams, and the resulting model is therefore a consistent extension of standard quantum mechanics, QM. This leads to elegant resolutions of single-particle interference paradoxes, but perhaps more important is how it handles entangled states. As is well known, Bell's theorem does not hold in MWI. As Lev Vaidman has put it, "in the framework of the MWI, Bell's argument cannot get off the ground because it requires a predetermined single outcome of a quantum experiment."[7] Although SI is a single-world and weaker version of MWI, it turns out that Bell's theorem also fails in SI, but for fundamentally different reasons. Indeed since the shadow system is consistent with standard quantum theory, it follows that it violates Bell's inequalities. Thus it only remains to show that local causality holds, which is done in Sec. IV in a rather simple and straightforward way.

N. David Mermin once wrote the following about attempts to find loopholes in Bell's theorem:

> In the intervening years [nearly two decades after the initial publication of his elegant and justly-celebrated conundrum version of Bell's theorem [8] I have found that the [conundrum's] transparency also makes it a good testing ground for claims of conceptual error in the formulation or proof of Bell's theorem. Confusion buried deep in the formalism of very general critiques tends to rise to the surface and reveal itself when such critiques are reduced to the language of my very elementary example. [9]

Our counterexample argument to Bell's theorem (of course not the same as identifying a gap in Bell's argument, though it has the same impact on the theorem's validity) is elementary, clean, and easy to check—nothing buried in the formalism. In fact it is the sort of thing that high-school students could understand completely once they were given some background in the elements of vector spaces and the inner-product operation.

Combining our result with the argument of BGHZ it then follows rather remarkably that, apart from MWI, there are just two choices for quantum foundations: systems such as SI containing hypothetical "somethings" or systems possessing non-local causality (David Bohm's pilot-wave theory [10] is an example of a non-local theory which reproduces the results of quantum mechanics). There can be no fence-sitting. As the old saying goes, you pays your money and you takes your choice; but assume special relativity and your choices narrow greatly.

This paper focuses on one- and two-particle interferometry, where there are only a finite number of alternative paths available to each particle, tangible or shadow. But, as mentioned, the general SI system (which may encompass uncountably many paths for a particle traveling between two points) is implemented mathematically by Feynman path integrals.[6]

SI, in addition to its relation to MWI, also bears some resemblance to the "sum over histories" (path-integral) approach of Murray Gell-Mann and James Hartle [11] (related to the "consistent histories" approach of Robert Griffiths and Roland Omnes) [12]) in which reality is taken to be a coarse-grained "class" of compatible paths or histories that have the same events in common. Sukanya Sinha and Rafael Sorkin [13] used sum-over-histories to provide a framework for quantum mechanics that possesses local causality and yet violates Bell's inequality (although there does not seem to be much discussion of their paper in the literature). The shadow interpretation, although it employs path integrals in a fundamental way, differs from the Sinha-Sorkin approach in that it does not use what they call "two-way histories." A two-way history views a pair of particles as traveling, say, in opposite directions from a source to a pair of detectors and then, on reaching the detectors, returning to the source.

## II. SHADOW POSTULATES, SHADOW STREAMS

> "An important addition to the knowledge of polarization was made in 1816 by Augustin J. Fresnel and D.F.J. Arago, who summed up the results of a searching series of experiments in the following laws of the interference of polarized light … [leading] at once to the conclusion that the stream may be represented by a vector …"
> (*The Encyclopedia Britannica*, Eleventh Edition, 1910-11)

> Thirty-one years ago [1949], Dick Feynman told me about his "sum over histories" version of quantum mechanics. "The electron does anything it likes," he said. "It just goes in any direction at any speed, forward or backward in time, however it likes, and then you add up the amplitudes and it gives you the wave-function." I said to him, "You're crazy." But he wasn't. (Freeman Dyson) [14]

As stated in Sec. I, we will assume the negation of EPNT ("emptiness of paths not taken") and also Deutsch's informal account of the properties of tangible and shadow interaction. We will also present five further postulates for the shadow system, all of which will be consistent with standard QM. Essentially, as noted in Sec. I, the new postulates amount to a "particle-implementation" of Richard Feynman's path-integral formulation of quantum mechanics, associating a particle with each path that a tangible particle can take when it travels between two points ([6] and (2.5) below).

Before listing the postulates, however, we recall the following elementary rules regarding amplitudes (from Feynman's classic *Lectures on Physics*, [15]) which some of our postulates extend and apply to the shadow system, and which suffice for most of the computations in this paper:

(2.1) The probability *P* of a quantum event is the square of the absolute value of a complex number $\phi$ (the probability amplitude): $P = \phi^*\phi = |\phi|^2$.

(2.2) When an event can occur in alternative ways, the combined amplitude, $\phi$, is the sum of the amplitudes for each way considered separately, $\phi = \phi_1 + \phi_2$

(2.3) When a particle goes by some particular route, the amplitude for that route is the product of the amplitude to go part way with the amplitude to go the rest of the way.

(2.4) Given two non-interacting particles, "the amplitude that one particle will do one thing and the other one do something else is the product of the two amplitudes that the two particles would do the two things separately."

Feynman famously emphasized that these are just computational rules, and that "No one has found any machinery behind the law. No one can 'explain' any more than we have just 'explained.' No one will give you any deeper representation of the situation." Nevertheless, as we shall see in Sec. III, the hypothetical entities referred to by BGHZ as "somethings," and by Deutsch as "shadow particles," do indeed provide machinery behind the law for single-particle interference (which is the kind of interference phenomena Feynman had in mind). Moreover, as we show in Sec. IV, they also neutralize paradoxes associated with multi-particle interference.

**Postulate 1:** Whenever a single, tangible particle travels from one point to another and has a choice of paths that it could take, it will (randomly) take one of the possible paths, and distinct shadow particles will take the others. The probability that it arrives at a particular point is given by a Feynman path-integral or sum (see (2.5) below). The collection of all such particles, the tangible particle and its counterpart shadow particles, is called "the shadow stream" associated with the tangible particle (the term "stream" is taken from the prescient 1910 *Encyclopedia Britannica* article cited above).

Throughout this paper we only consider particles moving in one spatial dimension over time (extending the system to cover more degrees of freedom is straightforward). Fig. 2.1 below illustrates various possible shadow streams generated by a tangible particle and its accompanying shadow particles as they move over several infinitesimal time stages (the picture is the kind that the mathematician Jerome Keisler termed "an infinitesimal snapshot").[16] Over an infinitesimal time interval the particle can be considered free (not subject to a potential), so its path is approximately linear—a ray.

The stage $t_k$ to $t_k + \varepsilon$ shows five shadow streams, each consisting of a single ray. From $t_k + \varepsilon$ to $t_k + 2\varepsilon$ there are three streams, each containing five rays. From $t_k + 2\varepsilon$ to $t_k + 3\varepsilon$ there are two shadow streams, each with three rays.

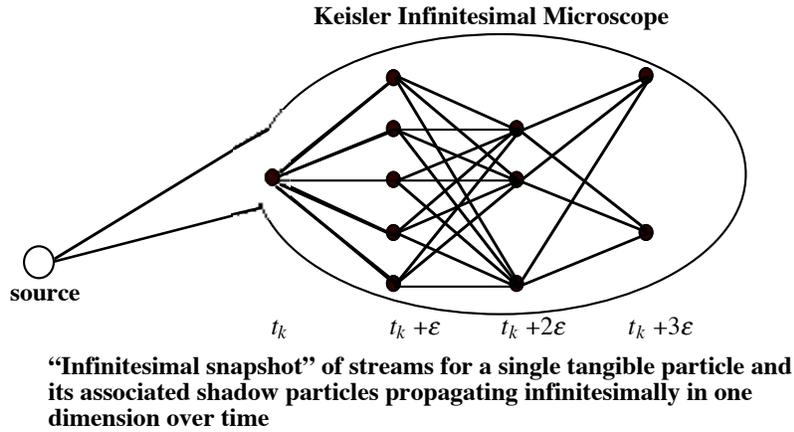

**Keisler Infinitesimal Microscope**

source — $t_k$ — $t_k + \varepsilon$ — $t_k + 2\varepsilon$ — $t_k + 3\varepsilon$

"Infinitesimal snapshot" of streams for a single tangible particle and its associated shadow particles propagating infinitesimally in one dimension over time

**Figure 2.1**

**Postulate 2** Distinct tangible particles generate distinct shadow streams.

**Postulate 3 (extending rule 2.2):** The amplitude of a tangible particle's shadow stream is the sum (integral) of the separate path-amplitudes for particles comprising the stream. This is what underlies single-particle interference. Notice, too, that the sum is only over path-amplitudes associated with particles in the stream. Amplitudes coming from different streams cannot be summed (see Postulate 4).

**Postulate 4 (extending rule 2.4):** By postulate 2 distinct tangible particles have distinct shadow streams. To quantify the combined amplitude for the separate shadow streams generated by distinct tangible particles we apply rule (2.4) above. Alternatively, one can take account of the fact that different streams can always be described by different Hilbert spaces in a tensor-product space, and so the combined amplitude is the multiple of the separate amplitudes (which of course is the same as a sum of multiples—associated with tensor products—of individual path amplitudes over the separate streams).

The next postulate formalizes a concept from Feynman's QED: [17] Consider a non-relativistic particle (tangible or shadow) starting from a point $a$ and moving along a path $\gamma(\tau)$. Let $S(\gamma(t)) = \int_0^t \left( \frac{m}{2}(\frac{d\gamma}{d\tau})^2 - V(\gamma(\tau)) \right) d\tau$ — the particle's action along the path $\gamma$. We then define the "path-clock" for that particle to be the time-dependent unit vector in the complex plane, $e^{iS(\gamma(t))/\hbar}$ (here $\hbar$ is Planck's constant reduced by $2\pi$, and $V$ is a potential). Thus when a non-relativistic particle (tangible or shadow) moves between two points, its path-clock begins rotating ("ticking") at a rate determined by its action. For a photon we simply assume that the clock rotates at the classical frequency of light.

As Ogborn and Taylor point out,

> Feynman's crucial and deep discovery was that you can base quantum mechanics on the postulate that [the difference between kinetic and potential energy] divided by the quantum of action $h$ is the rate of rotation of the quantum arrow. [18]

**Postulate 5 (path-clocks):** When a single particle (tangible or shadow) is emitted from a source its path-clock has a random initial value, and the clocks for the accompanying shadow particles are synchronized with the tangible particle's clock. When daughter particles are emitted from a source upon the decay of a mother particle, the clock values for the daughter particles have identical initial values—the common value being a random event. (The idea can be extended to cover any number of particles, but for the purposes of this paper we need to consider only two tangible daughter particles).

As noted, postulate 5 is essentially just an implementation of Feynman's "clock" associated with each particular, possible path for a tangible particle. But in the shadow system it implies a distinction that does not exist in the standard Hilbert-space formalism, which employs wavefunctions (kets, vectors) to describe tangible particles. In Hilbert-space formalism, any two wavefunctions that differ by a complex factor represent the same particle. Thus it is meaningless to talk about the phase of a particle (just multiply the wavefunction by $e^{i\theta}$ for an arbitrary $\theta$). In effect, although we will not make any deep use of this fact, postulate 5 implies a homomorphism from the shadow system (perhaps ultimately, from MWI—which was the provenance of Deutsch's original concept of shadow particles) onto the Hilbert space for position-momentum that describes our tangible universe.

SI is, in a way, a conceptual descendant of MWI. Thus it is not surprising that the shadow system can be extended to encompass spin: A similar construct (beyond the scope of the present paper) will encompass spin, but it is far less tractable mathematically: The "paths" for spin are in $\mathbb{R}^3 \times SO(3)$, where $SO(3)$ is the group manifold of rotations in Euclidean space, and where concentric, spinning shadow spheres, their radii differing infinitesimally, serve as the analogue of streams in coordinate space.

Consider now a single non-relativistic tangible particle, with its shadow stream propagating over an infinitesimal time interval as shown in Fig. 2.1 above. In this case the amplitude, $\psi(x,t)$, is given by a Feynman path sum (integral). The recursive expression for the (un-normalized) amplitude is, [6]

$$\psi(x,t+\varepsilon) = \sum_a \left( \exp[\frac{i}{\hbar} S(x,a)] \right) \psi(a,t). \qquad (2.5)$$

Here $S(x,a) = \int_t^{t+\varepsilon} \left( \frac{m}{2}(\frac{d\gamma}{d\tau})^2 - V(\gamma(\tau)) \right) d\tau$ is the action in going from $a$ to $x$ along the classical path $\gamma$, and $\varepsilon$ is infinitesimal.

In [6, 19] Feynman derived Schrödinger's wave equation from (2.5):

$$i\hbar \frac{\partial \psi}{\partial t} = -\frac{\hbar^2}{2m} \frac{\partial^2}{\partial x^2} \psi + V\psi$$

As can be seen from Fig. 2.1, a "classical" picture of particle interaction underlies the shadow system. Nevertheless, because of Feynman's derivation, **the shadow system is a consistent extension of standard quantum mechanics.[20]** When we deal with multi-particle systems in Sec. IV, this has the important consequence that violation of Bell's inequalities comes for free in the shadow system.

As originally developed, Feynman's approach seems to require that a particle must simultaneously explore all possible paths in traveling between two points (as in the above quote from Freeman Dyson; or as Feynman once put it in a different context, the particle

"smells all the paths in the neighborhood").[21] This is certainly a paradox, requiring a particle to move at superluminal speeds. In fact this suggestion of paradox may account for the formulation's being generally overlooked in introductory courses on quantum mechanics, which usually favor the Schrödinger differential-equation approach. In any case, Feynman's path-integral approach has considerable intuitive beauty and pedagogical merit, as E. F. Taylor has shown.[22, 23, 24]

The shadow interpretation eliminates the "explores all paths" paradox by associating a shadow particle with each possible path other than the one actually taken by the tangible particle (to reiterate, a random event—postulate 1). Moreover, as mentioned, the picture of particle interaction in SI is an easily visualizable, classical one—which contrasts with the standard doctrine. As N. David Mermin writes,

> It is a fundamental quantum doctrine that a measurement does not, in general, reveal a preexisting value of the measured property. On the contrary, the outcome of a measurement is brought into being by the act of measurement itself, a joint manifestation of the state of the probed system and the probing apparatus.[25]

Interestingly, in the shadow system we can view every particle in a shadow stream (Fig. 2.1) as having a definite position and momentum prior to measurement. Nevertheless, all the usual experimental predictions remain. Indeed, when we make a measurement in coordinate space, the basic procedures are the same, except for this difference: Instead of measuring the state of a particle we are actually measuring the state of the particle's shadow stream. Now, because of the properties of shadow and tangible interaction, our instruments can directly detect only observable properties pertaining to the *tangible* particle within the shadow stream. Any measurement we perform will nevertheless affect the shadow stream.

Take, for instance, the Heisenberg uncertainty principle. This principle is a valid mathematical consequence of representing the momentum wavefunction as the Fourier transform of the position wavefunction. But these wavefunctions describe the particle's shadow stream, not the particle itself. When we take a measurement to determine, say, the particle's position, we disturb the state of the particle's shadow stream (along the lines of what Bohr sometimes argued, but from a completely different viewpoint). This is because a wavefunction, apart from path-states, generally does not represent the particle's state, only the state of the corresponding shadow stream. By disturbing the shadow stream, any position measurement will have ramifications for the stream in which the particle is traveling, and therefore for the particle's momentum, and so forth. Textbooks on QM (in the spirit of the above doctrine) often claim that the particle does not have a simultaneous position and momentum. But of course they have no way of knowing or demonstrating this. Certainly it does not follow from the mathematics. What does follow, if we associate with each measurable (observable) quantity its set of eigenstates in an appropriate Hilbert space, is that we cannot "measure" the observables of position and momentum simultaneously. This is rather different from concluding that they do not exist simultaneously, for which—to repeat—there is no physical basis. (In Sec. IV we deal with Bell's theorem, which is often taken as evidence that such observables cannot exist simultaneously).[25]

The shadow stream makes physical sense of many otherwise puzzling quantum phenomena, imparting an almost classical meaning to them. Here are a couple of simple examples, starting with the expectation value of position for a stream in state $\psi$:

$$\langle x \rangle = \int_{-\infty}^{\infty} x |\psi(x,t)|^2 \, dx$$

In traditional approaches, $\langle x \rangle$ is interpreted as the average of the position measurements of an ensemble of tangible particles, each prepared in the same state $\psi$ —clearly a somewhat contorted notion. In contrast, for us it is simply the expected value for the particle's position in the stream at the time of measurement (the stream consisting of a tangible particle and its shadow counterparts).

Again, consider the time derivative of $\langle x \rangle$:

$$\frac{d}{dt}\langle x \rangle = \int_{-\infty}^{\infty} x \frac{\partial}{\partial t} |\psi(x,t)|^2 \, dx = -\frac{i\hbar}{m} \int_{-\infty}^{\infty} \psi^* \frac{\partial \psi}{\partial x} \, dx.$$

In the traditional approach this is the velocity of the expectation value of $x$, hardly a clear physical notion. For us, however, $\frac{d}{dt}\langle x \rangle$ has a straightforward meaning: The particles in the stream are moving over time, and $\frac{d}{dt}\langle x \rangle$ is the average velocity (change in position over time) of those particles at the instant of measurement.

Of course we could (if space allowed, for example in a textbook) similarly go on to employ streams to impart intuitive and classical significance to all of the usual paradigms of introductory QM, such as "particle-in-a-box," "particle-on-a-ring," and so forth.

## III. SINGLE-PARTICLE INTERFERENCE PARADOXES

The general SI system deals with streams containing uncountably many particles, but things are much simpler in an interferometer (our chief focus in this paper). In an interferometer there may be only a few possible paths from the source to a detector; and so the corresponding stream will contain a correspondingly small number of shadow particles. In this case Feynman's four rules in Sec. II, together with our postulates, suffice to quantify the experimental outcomes. Applying these, we see that the shadow system leads to rather elegant resolutions of various single-particle interference paradoxes. In this section we discuss a two examples that are based on interferometers, working with the common device of Fig. 3.1 below, which is one-half of the two-particle BGHZ device discussed in Sec. IV.

In the setup of Fig. 3.1 a single tangible particle enters the interferometer through the first beamsplitter and travels through to a detector. The device has two beamsplitters and two mirrors, and one of its arms contains a phase shifter. A straightforward calculation below shows that the probability for the tangible particle to arrive at detector $u$ is $\cos^2 \frac{\alpha}{2}$ and at $d$ is $\sin^2 \frac{\alpha}{2}$. Thus when $\alpha = 0$, the particle always ends up at $u$.

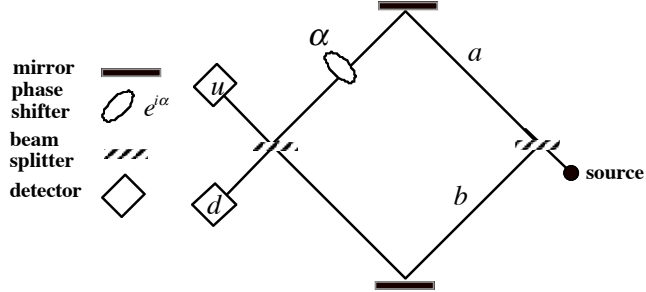

**Figure 3.1**

It will be instructive first to carry out the calculation in the conventional way using Dirac notation, and then do it from the point of view of the shadow stream and postulate 5.

In the conventional approach, we let $|u\rangle$ and $|d\rangle$ be an orthonormal basis for the two-dimensional vector space describing arrival at the detectors. We let $|\psi\rangle$ represent the initial state of the system, and let $|T\rangle$ and $|R\rangle$ be the transmission and reflection states after a beamsplitter (as usual because of unitarity[26] we attach an amplitude of $i$ for reflection at a beamsplitter):

$$|\psi\rangle = \frac{1}{\sqrt{2}}\left(e^{i\alpha}|T\rangle + i|R\rangle\right)$$

$$= \frac{1}{\sqrt{2}}\left(e^{i\alpha}\left(\frac{i|u\rangle+|d\rangle}{\sqrt{2}}\right) + i\left(\frac{|u\rangle+i|d\rangle}{\sqrt{2}}\right)\right)$$

$$= \frac{i(1+e^{i\alpha})}{2}|u\rangle + \frac{(-1+e^{i\alpha})}{2}|d\rangle$$

Thus

$$\langle u|\psi\rangle = \frac{i(1+e^{i\alpha})}{2} \tag{3.1}$$

$$(\langle u|\psi\rangle)^*\langle u|\psi\rangle = \frac{(1+e^{-i\alpha})}{2}\frac{(1+e^{i\alpha})}{2} = \cos^2\frac{\alpha}{2}$$

$$\langle d|\psi\rangle = \frac{(-1+e^{i\alpha})}{2} \tag{3.2}$$

$$(\langle d|\psi\rangle)^*\langle d|\psi\rangle = \frac{(-1+e^{-i\alpha})}{2}\frac{(-1+e^{i\alpha})}{2} = \sin^2\frac{\alpha}{2}$$

We will now analyze the forgoing calculation from the standpoint of the shadow stream and the path-clocks of postulate 5, which will be helpful when we look at the case of two entangled particles in Sec. IV.

There are two streams involved, corresponding to arrival at a detector. As before we let ket (i. e., vector) $|u\rangle$ represent arrival at detector $u$, and $|d\rangle$ arrival at detector $d$. Also we let ket $|a\rangle$ denote one of the particles traveling path $a$ (a tangible or shadow particle, but of course we don't know which), and let ket $|b\rangle$ denote the counterpart particle traveling path $b$. By postulate 5 when a particle leaves the first beamsplitter its clock is rotating. The three relevant clock values at the time it reaches the second beamsplitter are $\alpha$, $\pi/2$, and $\theta$—where $\theta$ is determined by the common path-length. We can then express each path-amplitude in terms of factors of these path-clocks. Of course the resulting stream amplitudes agree with those from the earlier calculation.

$$\text{amplitude to arrive at } u: \quad \overset{\text{path-amplitude:}\langle u|a\rangle}{\left(\frac{1}{2}e^{i(\theta+\alpha+\frac{\pi}{2})}\right)} + \overset{\text{path-amplitude:}\langle u|b\rangle}{\left(\frac{1}{2}e^{i(\frac{\pi}{2}+\theta)}\right)} = \frac{1}{2}e^{i\theta}i(e^{i\alpha}+1) \quad (3.3)$$

$$\text{amplitude to arrive at } d: \quad \overset{\text{path-amplitude:}\langle d|a\rangle}{\left(\frac{1}{2}e^{i\alpha}\right)} + \overset{\text{path-amplitude:}\langle d|b\rangle}{\left(\frac{1}{2}e^{i(\frac{\pi}{2}+\theta+\frac{\pi}{2})}\right)} = \frac{1}{2}e^{i\theta}(e^{i\alpha}-1) \quad (3.4)$$

We will now analyze Wheeler's famous "delayed choice paradox" from the shadow system perspective.

**Wheeler delayed-choice paradox.** In this paradox we suppose that a tangible photon is emitted from a source, passes through a beamsplitter, and then travels within a galactic-sized Mach-Zehnder interferometer to exit a second beamsplitter, eventually arriving at one of two detectors, $u$ or $d$. We can depict this setup by again using Fig. 3.1, but imagining now that the various distances in the apparatus are light years in length (Wheeler chose the astronomical scale to dramatize the resulting paradox).

Letting $\alpha = 0$ in the above calculation, we see that the tangible particle is always detected at $u$, never at $d$. But if we peek at the particle while it is in transit to locate in which arm of the interferometer it is actually traveling, even long after it has set out on its journey, we will destroy the interference, allowing the particle to reach the second detector, $d$, some of the time. In discussing this paradox the mathematical physicist Roger Penrose writes,

> "… the key puzzle is that somehow a photon (or other quantum particle) seems to have to 'know' what kind of experiment is going to be performed upon it well in advance of the actual performing of that experiment. How can it have the foresight to know whether to put itself into 'particle mode' or 'wave mode' as it leaves the (first) beamsplitter." [27]

Sometimes the paradox is also said to be an instance of "the future influencing the past" or "… a strange inversion of the normal order of time … an unavoidable effect on what we have a right to say about the already past history of that photon."[28]

R. B. Griffiths—one of the founders of "consistent histories," CH, mentioned in Sec. I—points out that "Wheeler used this paradox to argue, in effect, that the process by which the photon moves through the interferometer cannot be described or thought about in a coherent way. ... [Thus from this conventional standpoint] a quantum system is like a 'black box', and one should not try and figure out what is going on inside the box." Griffiths then goes on to say that the "whole field of quantum foundations is littered with these black boxes."[29]

Consistent histories is an important approach for resolving quantum paradoxes. Before presenting our simple resolution of the delayed-choice paradox we will, for contrast, briefly consider how CH peers inside the boxes.

A basic principle of the CH formalism (the "single-family framework rule") is that there are alternative frameworks or consistent families of histories (often infinitely many) associated with a quantum event, "all of which are considered 'equally valid' in the sense that no fundamental physical law determines which family should be used in any given case. … different frameworks are often mutually incompatible in a manner which means that the use of one to describe a given physical system precludes the use of another, that certain questions can only be addressed if one uses the appropriate framework, and that results from incompatible frameworks cannot be combined to form a single quantum description."[30]

This approach is obviously mathematically sound—a descendant of the von Neumann-Birkhoff quantum logic,[31, 32] but one that retains more of the feel of ordinary logic. Still, one must rigorously train one's intuition to apply the approach correctly—to appreciate "which questions can only be addressed if one uses the appropriate framework." This, in itself, is hardly a defect. But more serious for a system designed to eliminate paradoxes is that the system actually harbors its own paradoxical conception of reality, a consequence of the single-family framework rule:

> "… reality is such that it can be described in various alternative, incompatible ways, using descriptions which cannot be combined or compared."[33]

As some researchers have complained, "the fact that a mere syntactical rule, the single-family framework rule, can allow for two histories being separately true and yet render it nevertheless meaningless to consider them together, is rather puzzling. How can one accept this peculiar aspect of the formalism?"[34]

In contrast, the shadow approach is intuitively obvious and easy to comprehend, and also rather easy to assimilate (see also Sec. V, where we point that there are actually very few choices for acceptable interpretations of QM). Indeed, from the standpoint of the shadow interpretation the resolution of the Wheeler paradox is immediate. There is no need for quantum particles possessing foresight. The final result is accounted for in a purely mechanistic fashion involving inanimate objects. The tangible particle's path-clock, along with that of its shadow counterpart, are synchronized and rotating (postulate 5) when they begin their galactic journeys. The interaction between the tangible and shadow particles is purely in accordance with their initial programming: that is, they interact in accordance with "locally-programmed, action instructions" that all objects in the universe share—be they particles, balls, or stars (construing "action" in a way that makes sense not only for non-relativistic particles but also for photons). When the particles interact at key junctures (beamsplitters) they have a "memory" of their intergalactic travels stored in their clock values. The clock values determine where they go next. We emphasize that this is all local interaction. What makes possible the resolution of paradoxes is that the tangible and shadow particles, when they meet at a common juncture, exchange complementary information regarding their separate paths.

As a further illustration we will now consider interaction-free measurement.

**The Elitzur-Vaidman interaction-free measurement (IFM) paradox.[35]** In this setup we assume that the path lengths are adjusted so that the probability of the $u$ detector firing is 1 and that of $d$ firing is 0. Suppose further that the internal workings of the device are hidden from view and that our goal is to discover whether one of the arms is blocked by looking at the detection pattern at *u* and *d*. Elitzur and Vaidman state three possible outcomes when a particle is sent through the machine of Fig. 3.2 where, say, path *a* is blocked: (i) no detector clicks, (ii) detector *u* clicks, (iii) detector *d* clicks.

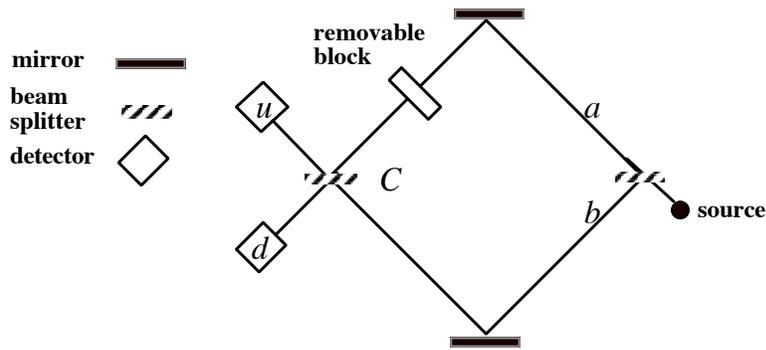

**Figure 3.2**

They point out that in the first case the probability for the outcome is 1/2, while in the second and the third case the probability is 1/4 because of the absence of interference. In the third case, they go on to claim: "We know that there is an object inside the interferometer without having 'touched' the object." This third outcome (which, to reiterate, happens 25% of the time) has been called "quantum seeing in the dark."[36] Clearly, it is remarkable. From our viewpoint, however, it might be called more aptly, "seeing with shadow particles." It is certainly no longer a paradox.

The shadow system elegantly eliminates a host of single- particle paradoxes (and, as we shall see below, it does the same for multi-particle cases, though in the multi-particle context the particle interaction is described by multiplication rather than addition of unit vectors).

# IV. HOW FEYNMAN'S CLOCKS CORRELATE BERTLMANN'S SOCKS[37]

> Indeed the hidden-variables theories ruled out by Bell's Theorem rest on assumptions that not only can be stated in entirely nontechnical terms but are so compelling that the establishment of their falsity has been called [38], not frivolously, " the most profound discovery of science."[39]

> Anybody who's not bothered by Bell's theorem has to have rocks in his head
> — Arthur Wightman, major contributor to quantum field theory [40]

> Most striking is the case of entanglement, which Einstein called "spooky," as it implies that the act of measuring a property of one particle can instantaneously change the state of another particle no matter how far apart the two are … . John Bell showed that the quantum predictions for entanglement are in conflict with local realism. From that "natural" point of view [local realism] any property we observe is (a) evidence of elements of reality out there and (b) independent of any actions taken at distant locations simultaneously with the measurement. Most physicists view the experimental confirmation of the quantum predictions as evidence for nonlocality. But I think that the concept of reality itself is at stake … .[41]

Truly, something important is at stake; and in this section we will look closely at BGHZ's "Bell theorem," showing how the denial of EPNT results in a system that possesses local causality and yet violates Bell's inequalities.

At the outset we note that violation of Bell's inequalities must occur in the shadow system, since the system is a consistent extension of standard quantum theory (because of Feynman's path-integral formulation). Indeed, the possibility is recognized by BGHZ when they write, "a something that samples the empty beam will still yield results in violation of the Bell inequality." In any case, it only remains to show that our system achieves its effects on the basis of local causality.

To this end, consider the following interferometer arrangement from BGHZ (Fig. 4.1). In this setup two tangible particles, 1 and 2, emerge from the source, S, and move along the paths a-a´ or b-b´. Each particle passes through a beamsplitter on its way to a detector. There are also two phase-shifters—$\alpha$ on path a, and $\beta$ on path b´.

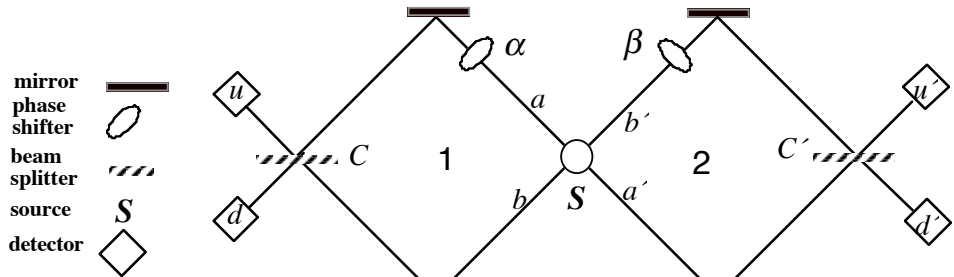

**Figure 4.1**

From a sequence of experiments in which beamsplitters are left in place or removed, BGHZ establish that the two tangible particles always (to repeat the point just noted above) take path a-a´ or b-b´. We will invoke this fact several times in what follows.

BGHZ let ket $|a\rangle_1$ denote particle 1 on path a, and so forth. Then BGHZ represent the initial quantum state of the pair of entangled tangible particles by

$$|\psi_0\rangle = \frac{1}{\sqrt{2}}\left(|a\rangle_1|a'\rangle_2 + |b\rangle_1|b'\rangle_2\right). \tag{4.1}$$

At the detectors the state equation (3.1) evolves into

$$|\psi\rangle = \frac{ie^{k(\alpha+\beta)}}{\sqrt{2}}\left[\left(|u\rangle_1|u'\rangle_2 + |d\rangle_1|d'\rangle_2\right)\cos(\frac{\alpha-\beta}{2}) + \left(|u\rangle_1|d'\rangle_2 - |d\rangle_1|u'\rangle_2\right)\sin(\frac{\alpha-\beta}{2})\right]. \tag{4.2}$$

Adjusting the phase shifters so that $\alpha = \beta$, produces only u-u´ and d-d´ coincidences at the detectors (half the time for each)—as predicted by equation (4.2). BGHZ calls this "perfect correlation," which they note is accounted for by the EPR program. But, as they point out, Bell's theorem does not apply to perfect correlation, only to "imperfect correlation"—i.e., where $\alpha$ and $\beta$ are allowed to vary arbitrarily. In this case, assuming EPNT, they show easily that EPR fails to account for imperfect correlations. So let us consider what happens when we negate EPNT.

BGHZ state the following (quoted earlier in Sec. I):

> … *something* could travel down the empty beam, so as to provide information to the nonempty beam, when the two beams meet. *And this something could be consistent with EPR locality, if the particles (and these somethings) on opposite sides of the origin do not communicate* [our emphasis]. However, one should be aware that there must be significant difference in the nature of this something in the one- and two-particle cases.

Here BGHZ are employing two terms that are pervasive in the lexicon of the Bell's-theorem literature—namely, "information" and "communicate," and (as vague as they may be in a technical sense) we also will find it convenient to use them in our discussion. But we emphasize that the kind of communication or sharing of information between particles that we have in mind is purely of a local, mechanical kind—physically exhibited by local wave-interference effects arising from the intersection of particle-paths at various points along the particles' travels (Fig. 2.1). As we did in Sec. III with Wheeler's paradox, we can perhaps best explain the underlying idea of our approach in terms of Feynman's clocks (postulate 5). Thus we will again view each particle (tangible or shadow) as being associated with a path-clock that begins rotating when the particle sets out from the source along its particular path. For a photon, the rate of rotation is determined by the classical frequency of light, and for a non-relativistic particle by the particle's action. When the tangible and shadow particles come together at, say, a beamsplitter, their path-clocks have a memory of the respective paths traveled to that point—that is, the clock values contain all the information needed to determine where the particles go next. The process is described mathematically by the addition of unit vectors when the particles come from the same stream, and by multiplication of vectors when they come from different streams (postulates 3 and 4).

Now consider what takes place when a mother particle decays to create a pair of tangible daughter-particles.

As BGHZ have established, the daughter tangible-particles will take paths *a-a′* or *b-b′*. Generally, the position of the source will vary randomly from trial to trial upon decay of the mother particle, but we will assume that the angle subtended by the paths *a* and *b′* with respect to the source is sufficiently small that there are just two choices for the upper tangible particle emitted from the source, either path *a* or path *b′*. This assumption ensures that the tangible and shadow particles taking paths *a* and *b′* will belong to the same stream (postulate 1), while the two taking *b* and *a′* belong to their own separate stream. This means that the particles taking paths *a* and *b* are in different streams. Similarly, for those taking *a′* and *b′*. By postulates 2 and 4 this ensures two-particle interference, described mathematically by multiplication of path-states in the tensor-product space ([42] presents the conventional explanation for two-particle interference in terms of the size of the angle subtended by the paths *a* and *b′*).

Suppose, without loss of generality, that one tangible particle takes path *a* and the other takes path *a′*. The shadow counterparts then take paths *b′* and *b*. We will let *t* and *s* stand for "tangible" and "shadow," with a subscript indicating which path the particle takes. Thus, for example, the tangible particle taking path *a* is $t_a$ and its shadow counterpart is $s_{b'}$, since the shadow counterpart takes path *b´* by postulate 1.

To recapitulate, the source event creates four particles and two streams. Thus $t_a$ and $s_{b'}$ are in the same shadow stream by postulate 1, as are $s_b$ and $t_{a'}$. But $t_a$ and $s_b$ are in different streams; so, too, are $s_{b'}$ and $t_{a'}$.

Nevertheless, when $t_a$ and $s_b$ meet at a beamsplitter their clocks contain complementary information about each particle's path through the left-half of the interferometer; similarly for $s_{b'}$ and $t_{a'}$ regarding paths on the right-side. Because $t_a$ and $s_b$ come from different streams, the shared clock information corresponds to multiplication of unit vectors in the associated path-states (postulate 4). Similarly for $s_{b'}$ and $t_{a'}$.

So far, however, all particle interaction is confined to each side separately; and yet, as countless experiments bear out, there is correlation between the outcomes on the two sides. In the (sadly) anthropomorphic mode one often encounters in such discussions, the claim is invariably made that for correlation to occur there must somehow be instantaneous communication (spooky) across the origin. But notice that the information carried by $s_b$ and $t_{a'}$ (stored in their respective clock values) is the same, since they take congruent paths, although the particles are confined to opposite sides. Nevertheless, on each side of the interferometer, when the separate particles encounter a beamsplitter, where they go next is determined by these clock values, postulates 4 and 5. But here is the key property of the shadow system that leads to correlation: The information stored in the associated clock values or path-amplitudes of $s_b$ and $t_{a'}$ is the same.

Let us now look at how all of this plays out mathematically.

Let $\langle u, u' | \psi_0 \rangle$ be the amplitude that the tangible-daughter pair arrives at detectors *u* and *u´*, and let $\langle u, d' | \psi_0 \rangle$ be the corresponding amplitude for arrival at *u* and *d´*. In all, there are four possible amplitudes for arrival at the detectors. The other two are $\langle d, u' | \psi_0 \rangle$, and $\langle d, d' | \psi_0 \rangle$). Without loss of generality, we will content ourselves with just calculating $\langle u, u' | \psi_0 \rangle$, and $\langle u, d' | \psi_0 \rangle$.

Working in the tensor-product space, we see that the amplitude $\langle u, u' | \psi_0 \rangle$ is the sum of two amplitudes, $\langle u|a\rangle\langle u'|a'\rangle$, and $\langle u|b\rangle\langle u'|b'\rangle$. This is the standard result, and it makes sense from our perspective of physical realism because, as BGHZ have established, the pair of daughter tangible-particles travel either the paths *a-a´* or the paths *b-b´*. Moreover, each of the summands is a product by postulates 3 and 4 (tangible and shadow particles coming from different streams), and also because that is how amplitudes are calculated in the tensor-product space (or to see this one can apply Feynman's rule (2.4) in Sec. II).

In Sec. III we showed in detail (equations (3.3) and (3.4)) how such a calculation corresponds to addition and multiplication of unit vectors in the complex plane. Here we simply present the outcomes:

$$\text{amplitude for } u\text{-}u' : \langle u, u' | \psi_0 \rangle$$
$$= \langle u|a\rangle\langle u'|a'\rangle + \langle u|b\rangle\langle u'|b'\rangle$$

a to u  a′ to u′  b to u  b′ to u′

$$= \left(\frac{1}{\sqrt{2}}e^{i\alpha}i\right)\left(\frac{1}{\sqrt{2}}\right) + \left(\frac{1}{\sqrt{2}}\right)\left(\frac{e^{i\beta}}{\sqrt{2}}i\right) = \frac{i}{2}(e^{i\alpha} + e^{i\beta}) \tag{4.3}$$

We have a similar result for $\langle u, d' | \psi_0 \rangle$:

$$\text{amplitude for } u\text{-}d' : \langle u, d' | \psi_0 \rangle$$
$$= \langle u|a\rangle\langle d'|a'\rangle + \langle u|b\rangle\langle d'|b'\rangle$$

a to u  a′ to d′  b to u  b′ to d′

$$= \left(\frac{1}{\sqrt{2}}e^{i\alpha}i\right)\left(\frac{1}{\sqrt{2}}i\right) + \left(\frac{1}{\sqrt{2}}\right)\left(\frac{e^{i\beta}}{\sqrt{2}}\right) = \frac{1}{2}(-e^{i\alpha} + e^{i\beta}) \tag{4.4}$$

Equations (4.3) and (4.4) yield the standard probability outcomes:

$$\langle u, u' | \psi_0 \rangle^* \langle u, u' | \psi_0 \rangle = \frac{1}{4}(e^{-i\alpha} + e^{-i\beta})(e^{i\alpha} + e^{i\beta})$$
$$= \frac{1}{4}\left(2 + e^{-i(\beta-\alpha)} + e^{i(\beta-\alpha)}\right) = \frac{1}{2}\cos^2\frac{\beta-\alpha}{2} \tag{4.5}$$

$$\langle u, d' | \psi_0 \rangle^* \langle u, d' | \psi_0 \rangle = \frac{1}{4}(-e^{-i\alpha} + e^{-i\beta})(-e^{i\alpha} + e^{i\beta})$$
$$= \frac{1}{4}\left(2 - e^{-i(\beta-\alpha)} - e^{i(\beta-\alpha)}\right) = \frac{1}{2}\sin^2\frac{\beta-\alpha}{2} \tag{4.6}$$

To reiterate an earlier point, on the face of it the equations leading to (4.3) and (4.4) have the appearance of non-local causality: For how can a particle moving along paths within one side of the interferometer coordinate its travels with a particle moving on the other side? As the usual story goes (incorrectly), there must be instantaneous communication across arbitrarily separated spatial regions: "How can these spatially separated locations 'know' what happens elsewhere? Is it all predetermined at the source? Do they somehow communicate? … that it is all predetermined at the source, has been ruled out by numerous Bell tests."[43]

Of course the appearance of nonlocality in itself proves nothing. For much the same reason—despite the assertion in the forgoing quote—no experiment (or at least none of the kind described so far in the literature) can rule out local causality. But here is where Bell's theorem is taken for granted. We have just seen that the inequalities are violated in our computation. So it seems to follow from Bell's theorem that local causality must fail. But wait. This is where shadow particles make all the difference.

As pointed out above, the shadow particle $s_b$ carries the same information as the tangible particle $t_{a'}$. The accumulated information is stored in each particle's associated path-clock as the particle travels through the interferometer. The information is the same for the simple reason that the particles take congruent paths. Thus,

$$\langle u|b\rangle = \langle u'|a'\rangle \text{ and } \langle d|b\rangle = \langle d'|a'\rangle. \tag{4.7}$$

Therefore making the substitutions from (4.7) into (4.3) and (4.4) we have,

$$\langle u|a\rangle\langle u'|a'\rangle + \langle u|b\rangle\langle u'|b'\rangle = \langle u|a\rangle\langle u|b\rangle + \langle u'|a'\rangle\langle u'|b'\rangle \tag{4.8}$$

$$\langle u|a\rangle\langle d'|a'\rangle + \langle u|b\rangle\langle d'|b'\rangle = \langle u|a\rangle\langle d|b\rangle + \langle u'|a'\rangle\langle d'|b'\rangle \tag{4.9}$$

As we see, in the second half of each of the equations (4.8) and (4.9) the information exchanged is purely local in nature when the tangible and shadow particles interact at the beamsplitters.

By postulates 4 and 5, this is the same kind of exchange as occurs at a beamsplitter in single-particle interference, although the exchange is associated with the multiplication of unit vectors (path-clocks) here rather than addition of vectors as in single-particle interference. In the present case, although the interaction of tangible and shadow particles at a beamsplitter has a random component (postulate 1), it determines—in each separate side of the device— which detector is reached by the tangible particle, and that in turn determines which detectors are reached by the pair of tangible particles traveling on opposite sides of the interferometer.

There is an interesting similarity here between particle interaction in the shadow system and what occurs in Brownian motion. As is well-known [44], there are striking parallels between the mathematics of the Wiener and Feynman integrals (as in the Einstein-Smoluchowski-Kolmogorov-Chapman relation, where transition probabilities correspond to transition amplitudes). In Brownian motion the process is Markovian (no "memory": the next state depends only on the present state, independently of past states). This holds also in the shadow system at intersection points such as at beamsplitters. Otherwise, however, there is a path-memory, stored in the path-amplitudes.

In any case, we see that each separate particle operates under purely local influences, going where the shared information between the tangible and shadow particles directs it. There is no communication across the origin, although the end-result is as though there were.

**QED** (but not in the sense that the mathematician John Allen Paulos [45] has somewhat liberally translated QED: "stick it in your eye.")

## V. CONCLUSION

> In my opinion, it is time to update the many quantum textbooks that introduce wavefunction collapse as a fundamental postulate of quantum mechanics … . If you are considering a quantum textbook that does not mention Everett [the many-worlds interpretation] and decoherence, I recommend buying a more modern one. (Max Tegmark)[46]

The physical significance of the result in Sec. IV is interesting. The BGHZ argument can be viewed as a carefully reasoned and detailed proof of the following implication:

$$\text{EPNT} \Rightarrow \text{non-local causality.}$$

But our above argument shows, both conceptually and mathematically,

$$\text{not-EPNT} \Rightarrow \text{local causality.}$$

We see that there are just two viable choices for our fundamental conception of quantum reality, no room for any more. And then, adding special relativity and Lorentz invariance (confirmed in countless experiments), there would be only one—as in the old nursery rhyme. This conclusion is, among other things, a testament to the supreme physical insight of Einstein and Everett (see also the comment in the discussion following postulate 5). Indeed, it appears that Tegmark in the above quote was actually entitled to make a much stronger recommendation, à la Hume.

---

[1] H. J. Bernstein, D. M. Greenberger, M. A. Horne, A. Zeilinger, "Bell theorem without inequalities for two spinless particles," Physical Review A, Vol. 47, Number 1, January, 1993

[2] D. Deutsch, *The Fabric of Reality*, Penguin Press, 1998

[3] J. G. Rarity, P. R. Tapster, "Experimental Violation of Bell's Inequality Based on Phase and Momentum," Physical Review Letters, Vol. 65, Number 21, 21 May 1990

[4] A. Einstein, B. Podolsky, N. Rosen, "Can quantum mechanical description of reality be considered complete?" Phys. Rev. 47, 777 (1935).

[5] J. A. Wheeler, "Law Without Law" in Problems in the Foundation of Physics, Proceedings of the International School of Physics 'Enrico Fermi,' Varenna on Lake Como 25th July-6th August 1977, North-Holland Publishing Company, 1979, p. 396.

[6] R.P. Feynman, "Space-Time Approach to Non-Relativistic Quantum Mechanics," Rev. Mod. Phys. 20, 367-387 (1948) reprinted in *Feynman's Thesis, A New Approach to Quantum Mechanics*, edited by Laurie M. Brown, World Scientific Publishing, 2005.

[7] L. Vaidman, "Many Worlds Interpretation of Quantum Mechanics," Stanford Encyclopedia of Philosophy, 2002

[8] N. D. Mermin, "Bringing home the quantum world: Quantum mysteries for anybody," Am. J. Phys. 49[10, Oct. 1981]

[9] N. David Mermin, "Shedding (red and green) light on 'time related hidden parameters'" arXiv:quant-ph/020618v1 18 Jun 2002

[10] D. Bohm, "A Suggested Interpretation of the Quantum Theory in Terms of 'Hidden Variables'," 1952, Physical Review **85**: 180–193.

[11] J. B. Hartle, M. Gell-Mann, "Quasiclassical coarse graining and thermodynamic entropy," Physical Review A 76, (2007)

[12] R.B. Griffiths, Roland Omnès. "Consistent Histories and Quantum Measurements," *Physics Today*, 52, (8), Part1. pp. 26-31, 1999

[13] S. Sinha, R. Sorkin, "A Sum-Over-Histories Account of an EPR (B) Experiment," Found. Phys. Lett. 4: 303-335 (1991)